\documentclass[10pt,twocolumn]{article}

\usepackage[utf8]{inputenc}
\usepackage[T1]{fontenc}
\usepackage{amsmath,amssymb}
\usepackage{graphicx}
\usepackage{booktabs}
\usepackage{algorithm}
\usepackage{algpseudocode}
\usepackage{hyperref}
\usepackage{url}
\usepackage{xcolor}
\usepackage{listings}
\usepackage{caption}
\usepackage{subcaption}
\usepackage{multirow}
\usepackage[margin=0.75in]{geometry}
\usepackage{enumitem}
\usepackage{balance}

\usepackage{tikz}
\usetikzlibrary{arrows.meta, positioning, shapes.geometric, fit, calc, decorations.pathreplacing, backgrounds, patterns}

\newcommand{\architectureDiagram}{
\begin{figure*}[t]
  \centering
  \begin{tikzpicture}[
    >=Stealth,
    phase/.style={draw, rounded corners=3pt, minimum width=3.2cm, minimum height=1.4cm, align=center, font=\small\bfseries},
    subitem/.style={font=\scriptsize, text=black!70, align=left},
    io/.style={draw, rounded corners=6pt, minimum width=2.4cm, minimum height=0.9cm, align=center, font=\small, fill=black!5, dashed},
    arrow/.style={->, thick, black!70},
    label/.style={font=\scriptsize\itshape, text=black!50},
  ]
    \node[io] (input) {Satellite ID\\Workload DAG};

    \node[phase, fill=blue!8, right=1.2cm of input] (p1) {Phase 1\\Environment};
    \node[subitem, below=0.1cm of p1] (p1s) {SGP4 propagation\\Eclipse detection\\Pass prediction\\Link budgets};

    \node[phase, fill=green!8, right=1.0cm of p1] (p2) {Phase 2\\Placement};
    \node[subitem, below=0.1cm of p2] (p2s) {Topological sort\\Reduction heuristic\\Cost comparison};

    \node[phase, fill=orange!8, right=1.0cm of p2] (p3) {Phase 3\\Transfers};
    \node[subitem, below=0.1cm of p3] (p3s) {Boundary detection\\Adaptive FEC\\Security overhead\\Multi-pass alloc.};

    \node[phase, fill=red!8, right=1.0cm of p3] (p4) {Phase 4\\Scheduling};
    \node[subitem, below=0.1cm of p4] (p4s) {Window assignment\\Resource checks\\Deadline enforce.};

    \node[io, right=1.2cm of p4] (output) {Execution\\Plan};

    \draw[arrow] (input) -- (p1);
    \draw[arrow] (p1) -- (p2);
    \draw[arrow] (p2) -- (p3);
    \draw[arrow] (p3) -- (p4);
    \draw[arrow] (p4) -- (output);

    \draw[arrow, dashed, blue!40] (p1.south) -- ++(0,-1.8) -| (p3.south);
    \draw[arrow, dashed, blue!40] (p1.south) -- ++(0,-1.8) -| (p4.south);
    \node[label] at ($(p1.south)+(3.5,-1.95)$) {environment data};

  \end{tikzpicture}
  \caption{CAE four-phase planning pipeline. The orbital environment computed in Phase 1 is consumed by all subsequent phases. Each phase refines the plan incrementally: placement decides where steps run, transfer insertion adds data movement steps, and scheduling assigns everything to orbital windows.}
  \label{fig:architecture}
\end{figure*}
}

\newcommand{\workloadDiagram}{
\begin{figure}[t]
  \centering
  \begin{tikzpicture}[
    >=Stealth,
    onboard/.style={draw, rounded corners=2pt, fill=cyan!12, minimum width=2.0cm, minimum height=0.6cm, align=center, font=\scriptsize},
    ground/.style={draw, rounded corners=2pt, fill=yellow!15, minimum width=2.0cm, minimum height=0.6cm, align=center, font=\scriptsize},
    transfer/.style={draw, rounded corners=2pt, fill=orange!15, minimum width=2.0cm, minimum height=0.6cm, align=center, font=\scriptsize, dashed},
    data/.style={font=\tiny, text=black!60},
    arrow/.style={->, thick, black!60},
  ]
    \node[onboard] (cap) at (0, 0) {Capture};
    \node[onboard] (feat) at (0, -1.1) {Feature Extract};
    \node[onboard] (comp) at (0, -2.2) {Compress};
    \node[onboard] (enc1) at (0, -3.3) {Encrypt};

    \node[transfer] (dl) at (0, -4.4) {Downlink};

    \node[ground] (train) at (0, -5.5) {Train Backend};
    \node[ground] (comp2) at (0, -6.6) {Compress Weights};
    \node[ground] (enc2) at (0, -7.7) {Encrypt Weights};

    \node[transfer] (ul) at (0, -8.8) {Uplink};

    \node[onboard] (deploy) at (0, -9.9) {Deploy Model};

    \draw[arrow] (cap) -- (feat);
    \draw[arrow] (feat) -- (comp);
    \draw[arrow] (comp) -- (enc1);
    \draw[arrow] (enc1) -- (dl);
    \draw[arrow] (dl) -- (train);
    \draw[arrow] (train) -- (comp2);
    \draw[arrow] (comp2) -- (enc2);
    \draw[arrow] (enc2) -- (ul);
    \draw[arrow] (ul) -- (deploy);

    \node[data, right=0.3cm of cap] {2000 MB};
    \node[data, right=0.3cm of feat] {50 MB (40:1)};
    \node[data, right=0.3cm of comp] {35 MB};
    \node[data, right=0.3cm of enc1] {36.75 MB};
    \node[data, right=0.3cm of dl] {52.8 MB w/ FEC};
    \node[data, right=0.3cm of train] {8 MB};
    \node[data, right=0.3cm of comp2] {5 MB};
    \node[data, right=0.3cm of enc2] {5.25 MB};
    \node[data, right=0.3cm of ul] {7.6 MB w/ FEC};

    \node[font=\scriptsize\bfseries, text=cyan!60!black, rotate=90] at (-1.8, -1.65) {ON-BOARD};
    \node[font=\scriptsize\bfseries, text=yellow!60!black, rotate=90] at (-1.8, -6.6) {GROUND};
    \node[font=\scriptsize\bfseries, text=cyan!60!black] at (-1.8, -9.9) {\rotatebox{90}{OB}};

    \draw[densely dotted, thick, orange!60] (-2.3, -3.85) -- (3.0, -3.85);
    \draw[densely dotted, thick, orange!60] (-2.3, -8.25) -- (3.0, -8.25);
    \node[font=\tiny\itshape, text=orange!70!black] at (3.6, -3.85) {boundary};
    \node[font=\tiny\itshape, text=orange!70!black] at (3.6, -8.25) {boundary};

    \node[onboard, minimum width=0.8cm] at (-1.2, -11.0) {};
    \node[font=\tiny, right] at (-0.6, -11.0) {On-board};
    \node[ground, minimum width=0.8cm] at (0.9, -11.0) {};
    \node[font=\tiny, right] at (1.5, -11.0) {Ground};
    \node[transfer, minimum width=0.8cm] at (2.7, -11.0) {};
    \node[font=\tiny, right] at (3.3, -11.0) {Transfer};

  \end{tikzpicture}
  \caption{Split learning workload DAG after placement and transfer insertion. Feature extraction achieves 40:1 data reduction on-board, minimizing downlink volume. Transfer steps (dashed) are automatically inserted at space-ground boundaries by Phase 3. Data volumes include FEC and security overhead.}
  \label{fig:workload_dag}
\end{figure}
}

\newcommand{\timelineDiagram}{
\begin{figure*}[t]
  \centering
  \begin{tikzpicture}[
    >=Stealth,
    font=\scriptsize,
  ]
    \def\tlen{15}
    \draw[->, thick] (0,0) -- (\tlen+0.5, 0) node[right] {Time};

    \foreach \x/\label in {0/0, 2.78/30, 5.56/60, 8.34/90, 11.12/120, 13.9/150} {
      \draw (\x, -0.1) -- (\x, 0.1);
      \node[below, font=\tiny] at (\x, -0.15) {\label min};
    }

    \def\ry{0.8}
    \node[left, font=\scriptsize\bfseries] at (-0.1, \ry) {Power};
    \fill[yellow!20] (0, \ry-0.25) rectangle (1.8, \ry+0.25);
    \fill[blue!10] (1.8, \ry-0.25) rectangle (3.2, \ry+0.25);
    \fill[yellow!20] (3.2, \ry-0.25) rectangle (5.0, \ry+0.25);
    \fill[blue!10] (5.0, \ry-0.25) rectangle (6.4, \ry+0.25);
    \fill[yellow!20] (6.4, \ry-0.25) rectangle (8.2, \ry+0.25);
    \fill[blue!10] (8.2, \ry-0.25) rectangle (9.6, \ry+0.25);
    \fill[yellow!20] (9.6, \ry-0.25) rectangle (11.4, \ry+0.25);
    \fill[blue!10] (11.4, \ry-0.25) rectangle (12.8, \ry+0.25);
    \fill[yellow!20] (12.8, \ry-0.25) rectangle (14.6, \ry+0.25);

    \node[font=\tiny] at (0.9, \ry) {80W};
    \node[font=\tiny] at (2.5, \ry) {25W};
    \node[font=\tiny, text=yellow!50!black] at (4.1, \ry+0.4) {sunlit};
    \node[font=\tiny, text=blue!50!black] at (5.7, \ry+0.4) {eclipse};

    \def\ryy{1.8}
    \node[left, font=\scriptsize\bfseries] at (-0.1, \ryy) {Passes};
    \fill[green!25] (1.0, \ryy-0.2) rectangle (1.6, \ryy+0.2);
    \node[font=\tiny, above] at (1.3, \ryy+0.2) {SVB};
    \fill[green!25] (4.2, \ryy-0.2) rectangle (5.1, \ryy+0.2);
    \node[font=\tiny, above] at (4.65, \ryy+0.2) {FAI};
    \fill[green!25] (7.8, \ryy-0.2) rectangle (8.5, \ryy+0.2);
    \node[font=\tiny, above] at (8.15, \ryy+0.2) {AWA};
    \fill[green!25] (11.0, \ryy-0.2) rectangle (11.5, \ryy+0.2);
    \node[font=\tiny, above] at (11.25, \ryy+0.2) {SNG};
    \fill[green!25] (13.5, \ryy-0.2) rectangle (14.2, \ryy+0.2);
    \node[font=\tiny, above] at (13.85, \ryy+0.2) {ORE};

    \def\ryyy{2.8}
    \node[left, font=\scriptsize\bfseries] at (-0.1, \ryyy) {Schedule};

    \fill[cyan!25, draw=cyan!50] (0.1, \ryyy-0.2) rectangle (0.9, \ryyy+0.2);
    \node[font=\tiny] at (0.5, \ryyy) {cap};
    \fill[cyan!25, draw=cyan!50] (3.3, \ryyy-0.2) rectangle (4.1, \ryyy+0.2);
    \node[font=\tiny] at (3.7, \ryyy) {feat};
    \fill[cyan!25, draw=cyan!50] (6.5, \ryyy-0.2) rectangle (7.0, \ryyy+0.2);
    \node[font=\tiny] at (6.75, \ryyy) {enc};

    \fill[orange!25, draw=orange!50] (7.8, \ryyy-0.2) rectangle (8.5, \ryyy+0.2);
    \node[font=\tiny] at (8.15, \ryyy) {DL};

    \fill[yellow!20, draw=yellow!60!black] (8.6, \ryyy+0.3) rectangle (10.0, \ryyy+0.7);
    \node[font=\tiny] at (9.3, \ryyy+0.5) {train (ground)};

    \fill[orange!25, draw=orange!50] (11.0, \ryyy-0.2) rectangle (11.5, \ryyy+0.2);
    \node[font=\tiny] at (11.25, \ryyy) {UL};

    \fill[cyan!25, draw=cyan!50] (12.9, \ryyy-0.2) rectangle (13.4, \ryyy+0.2);
    \node[font=\tiny] at (13.15, \ryyy) {dep};

    \draw[->, thin, black!40] (0.9, \ryyy) -- (3.3, \ryyy);
    \draw[->, thin, black!40] (4.1, \ryyy) -- (6.5, \ryyy);
    \draw[->, thin, black!40] (7.0, \ryyy) -- (7.8, \ryyy);
    \draw[->, thin, black!40] (8.5, \ryyy) -- (8.6, \ryyy+0.5);
    \draw[->, thin, black!40] (10.0, \ryyy+0.5) -- (11.0, \ryyy);
    \draw[->, thin, black!40] (11.5, \ryyy) -- (12.9, \ryyy);

    \fill[cyan!25, draw=cyan!50] (0, -0.9) rectangle (0.5, -0.6);
    \node[font=\tiny, right] at (0.55, -0.75) {On-board compute};
    \fill[orange!25, draw=orange!50] (3.0, -0.9) rectangle (3.5, -0.6);
    \node[font=\tiny, right] at (3.55, -0.75) {Data transfer};
    \fill[yellow!20, draw=yellow!60!black] (5.8, -0.9) rectangle (6.3, -0.6);
    \node[font=\tiny, right] at (6.35, -0.75) {Ground compute};
    \fill[green!25] (8.8, -0.9) rectangle (9.3, -0.6);
    \node[font=\tiny, right] at (9.35, -0.75) {Pass window};

  \end{tikzpicture}
  \caption{Example schedule for the split learning workload on the ISS. The top row shows power state (sunlit at 80W, eclipse at 25W). The middle row shows ground station pass windows. The bottom row shows the scheduled execution: on-board steps (cyan) are placed in sunlit windows for full compute capacity, transfers (orange) are pinned to pass windows, and ground compute (yellow) begins immediately after data arrives. Abbreviations: SVB = Svalbard, FAI = Fairbanks, AWA = Awarua, SNG = Singapore, ORE = Oregon.}
  \label{fig:timeline}
\end{figure*}
}

\newcommand{\overheadDiagram}{
\begin{figure}[t]
  \centering
  \begin{tikzpicture}[
    font=\scriptsize,
  ]
    \def\bh{0.45}  
    \def\gap{0.85} 

    \def\sc{0.012}

    \def\y{4*\gap}
    \node[left, font=\tiny, align=right] at (0, \y) {ML\\Inference};
    \fill[cyan!30] (0, \y-\bh/2) rectangle ({10.5*\sc}, \y+\bh/2);
    \fill[red!25] ({10.5*\sc}, \y-\bh/2) rectangle ({14.0*\sc}, \y+\bh/2);
    \fill[purple!20] ({14.0*\sc}, \y-\bh/2) rectangle ({14.7*\sc}, \y+\bh/2);
    \node[font=\tiny, right] at ({14.7*\sc}, \y) {14.7};

    \def\y{3*\gap}
    \node[left, font=\tiny, align=right] at (0, \y) {Split\\Learning};
    \fill[cyan!30] (0, \y-\bh/2) rectangle ({36.75*\sc}, \y+\bh/2);
    \fill[red!25] ({36.75*\sc}, \y-\bh/2) rectangle ({49.0*\sc}, \y+\bh/2);
    \fill[purple!20] ({49.0*\sc}, \y-\bh/2) rectangle ({52.8*\sc}, \y+\bh/2);
    \node[font=\tiny, right] at ({52.8*\sc}, \y) {52.8};

    \def\y{2*\gap}
    \node[left, font=\tiny, align=right] at (0, \y) {EO\\QA};
    \fill[cyan!30] (0, \y-\bh/2) rectangle ({400*\sc}, \y+\bh/2);
    \fill[red!25] ({400*\sc}, \y-\bh/2) rectangle ({533*\sc}, \y+\bh/2);
    \fill[purple!20] ({533*\sc}, \y-\bh/2) rectangle ({560*\sc}, \y+\bh/2);
    \node[font=\tiny, right] at ({560*\sc}, \y) {560};

    \def\y{1*\gap}
    \node[left, font=\tiny, align=right] at (0, \y) {Federated};
    \fill[cyan!30] (0, \y-\bh/2) rectangle ({3.5*\sc}, \y+\bh/2);
    \fill[red!25] ({3.5*\sc}, \y-\bh/2) rectangle ({4.67*\sc}, \y+\bh/2);
    \fill[purple!20] ({4.67*\sc}, \y-\bh/2) rectangle ({5.0*\sc}, \y+\bh/2);
    \node[font=\tiny, right] at ({5.0*\sc}, \y) {5.0};

    \def\y{0*\gap}
    \node[left, font=\tiny, align=right] at (0, \y) {Store-\\Forward};
    \fill[cyan!30] (0, \y-\bh/2) rectangle ({100*\sc}, \y+\bh/2);
    \fill[red!25] ({100*\sc}, \y-\bh/2) rectangle ({150*\sc}, \y+\bh/2);
    \fill[purple!20] ({150*\sc}, \y-\bh/2) rectangle ({157.5*\sc}, \y+\bh/2);
    \node[font=\tiny, right] at ({157.5*\sc}, \y) {157.5};

    \draw[->] (0, -0.5) -- ({600*\sc+0.2}, -0.5) node[right, font=\tiny] {MB};
    \foreach \x in {0, 200, 400, 600} {
      \draw ({\x*\sc}, -0.55) -- ({\x*\sc}, -0.45);
      \node[font=\tiny, below] at ({\x*\sc}, -0.55) {\x};
    }

    \fill[cyan!30] (0, -1.1) rectangle (0.3, -0.9);
    \node[font=\tiny, right] at (0.35, -1.0) {Raw};
    \fill[red!25] (1.3, -1.1) rectangle (1.6, -0.9);
    \node[font=\tiny, right] at (1.65, -1.0) {FEC};
    \fill[purple!20] (2.5, -1.1) rectangle (2.8, -0.9);
    \node[font=\tiny, right] at (2.85, -1.0) {Sec.+frame};

  \end{tikzpicture}
  \caption{Transfer volume breakdown (MB) by workload. Raw data (cyan) is the payload after on-board processing. FEC overhead (red) depends on channel quality. Security and CCSDS framing (purple) adds encryption and packet headers.}
  \label{fig:overhead}
\end{figure}
}

\newcommand{\reductionDiagram}{
\begin{figure}[b]
  \centering
  \begin{tikzpicture}[
    font=\scriptsize,
    bar/.style={draw=black!30, thick},
  ]

    \def\bw{0.8}

    \fill[cyan!40, bar] (0, 0) rectangle (\bw, 5.0);
    \node[font=\tiny, above] at (\bw/2, 5.0) {2000 MB};
    \node[font=\tiny, below] at (\bw/2, -0.15) {Capture};

    \fill[cyan!35, bar] (1.5, 0) rectangle (1.5+\bw, 3.5);
    \node[font=\tiny, above] at (1.5+\bw/2, 3.5) {500 MB};
    \node[font=\tiny, below] at (1.5+\bw/2, -0.15) {Preproc.};
    \draw[->, thick, red!60!black] (0.9, 4.2) -- (1.4, 3.7);
    \node[font=\tiny\bfseries, text=red!70!black] at (1.15, 4.25) {4:1};

    \fill[cyan!30, bar] (3.0, 0) rectangle (3.0+\bw, 0.5);
    \node[font=\tiny, above] at (3.0+\bw/2, 0.5) {10 MB};
    \node[font=\tiny, below] at (3.0+\bw/2, -0.15) {Inference};
    \draw[->, thick, red!60!black] (2.4, 3.0) -- (2.9, 0.8);
    \node[font=\tiny\bfseries, text=red!70!black] at (2.3, 2.2) {50:1};

    \fill[orange!30, bar] (4.5, 0) rectangle (4.5+\bw, 0.52);
    \node[font=\tiny, above] at (4.5+\bw/2, 0.52) {10.5 MB};
    \node[font=\tiny, below] at (4.5+\bw/2, -0.15) {Encrypt};

    \draw[<->, very thick, red!50!black] (6.0, 0) -- (6.0, 5.0);
    \node[font=\scriptsize\bfseries, text=red!60!black, rotate=90, align=center] at (6.5, 2.5) {190:1};

    \node[font=\tiny, rotate=90] at (-0.5, 2.5) {Data volume (MB)};

    \draw[dashed, black!30] (-0.3, 0.5) -- (5.5, 0.5);
    \node[font=\tiny, text=black!50, right] at (5.5, 0.5) {downlink};

  \end{tikzpicture}
  \caption{Data reduction cascade for the on-board ML inference workload. Each processing step reduces data volume, with inference achieving 50:1 compression. The overall 190:1 reduction means only 10.5 MB needs to be downlinked from the original 2 GB capture. Bar heights use a compressed scale for visibility.}
  \label{fig:reduction}
\end{figure}
}

\newcommand{\eclipseDiagram}{
\begin{figure}[t]
  \centering
  \begin{tikzpicture}[
    >=Stealth,
    font=\scriptsize,
  ]
    \shade[ball color=blue!20] (0,0) circle (1.2);
    \node[font=\tiny] at (0,0) {Earth};

    \fill[black!8] (-5, -1.2) rectangle (0, 1.2);
    \draw[dashed, black!30] (0, 1.2) -- (-5, 1.2);
    \draw[dashed, black!30] (0, -1.2) -- (-5, -1.2);
    \node[font=\tiny, text=black!50] at (-2.5, 0) {shadow cylinder};

    \draw[->, very thick, yellow!70!orange] (4, 0) -- (1.5, 0);
    \node[font=\small] at (4.5, 0) {$\odot$};
    \node[font=\tiny, below] at (4.5, -0.2) {Sun};

    \fill[green!70!black] (2.0, 1.5) circle (0.08);
    \node[font=\tiny, above] at (2.0, 1.6) {sat (sunlit)};
    \draw[thin, green!50!black] (2.0, 1.5) -- (0, 0);
    \node[font=\tiny, text=green!50!black] at (1.3, 1.15) {$d > 0$};

    \fill[red!70!black] (-2.5, 0.7) circle (0.08);
    \node[font=\tiny, above] at (-2.5, 0.8) {sat (eclipsed)};
    \draw[thin, red!50!black] (-2.5, 0.7) -- (-2.5, 0);
    \draw[<->, thin, red!50!black] (-2.5, 0) -- (-2.5, 0.65);
    \node[font=\tiny, text=red!50!black, right] at (-2.4, 0.35) {$r_\perp < R_\oplus$};

    \draw[thin, dotted] (-2.5, 0) -- (0, 0);

    \draw[<->, thin] (0.15, 0) -- (0.15, 1.2);
    \node[font=\tiny, right] at (0.2, 0.6) {$R_\oplus$};

  \end{tikzpicture}
  \caption{Cylindrical shadow model for eclipse detection. A satellite is in eclipse when it lies on the anti-sun side of Earth ($d < 0$) and its perpendicular distance from the Earth-Sun line is less than Earth's radius ($r_\perp < R_\oplus$). The model treats the shadow as an infinite cylinder, neglecting penumbra and atmospheric refraction.}
  \label{fig:eclipse}
\end{figure}
}

\title{Constraint-Aware Execution Planning for Hybrid Space-Ground Compute Workloads}

\author{
  Subhadip Mitra \\
  RotaStellar \\
  \texttt{subhadip@rotastellar.com}
}

\date{March 2026}

\begin{document}

\maketitle

\begin{abstract}
Low Earth orbit (LEO) satellites increasingly carry compute hardware capable of on-board processing, yet each satellite generates roughly two orders of magnitude more data than it can downlink per orbit. This mismatch forces operators to decide, for every workload, which computation runs on-board and which runs on the ground, how intermediate data crosses the space-ground boundary through narrow contact windows, and how to maintain delivery guarantees over noisy channels. We present Constraint-Aware Execution (CAE), a planning system that takes a satellite identifier, a workload expressed as a directed acyclic graph of processing steps, and a set of orbital and resource constraints, and produces a deterministic, physically grounded execution plan. CAE operates in four phases: (1) orbital environment construction via SGP4 propagation with eclipse detection and ground station pass prediction, (2) compute placement using a cost model that compares on-board resource consumption against transfer overhead, (3) transfer insertion with adaptive forward error correction and security overhead modeling, and (4) greedy first-fit scheduling into orbital windows under power, thermal, compute, and communication constraints. We evaluate CAE against five representative workload patterns across satellites in distinct orbital regimes and demonstrate that the system produces feasible plans in under two seconds, correctly exploits on-board data reduction to minimize transfer volume, and adapts FEC and multi-pass allocation to varying channel conditions. CAE is deployed as a production API computing plans for any cataloged satellite using live two-line element data.
\end{abstract}

\section{Introduction}
\label{sec:introduction}

A modern Earth observation satellite in low Earth orbit generates on the order of 1 terabyte of raw data per day~\cite{selva2012survey}. A typical 10-minute ground station pass using X-band communications can transfer approximately 7.5 gigabytes. This represents a roughly 130:1 mismatch between data generation and downlink capacity, and the ratio is worsening as sensor resolution increases while orbital mechanics constrain contact windows.

The traditional operational model - capture on-board, downlink everything, process on the ground - is reaching its limits. Simultaneously, on-board compute hardware has become substantially more capable. Radiation-tolerant GPUs, FPGAs, and custom accelerators now fly routinely on commercial satellites, enabling meaningful on-board processing~\cite{furano2020ai, giuffrida2021cloudscout}.

This creates a hybrid compute environment where some processing should happen on-board (particularly steps that dramatically reduce data volume before downlink) while other processing belongs on the ground (model training with large datasets, aggregation across satellite constellations, anything requiring persistent storage). The challenge lies at the boundary: scheduling data transfers across intermittent contact windows with variable link quality, managing error correction overhead, and producing plans that respect the physical constraints of orbital mechanics.

Existing approaches to this problem fall into three categories. Mission planning tools from traditional space agencies handle scheduling at the mission level but do not reason about compute placement~\cite{johnston2014automated}. Edge computing frameworks designed for terrestrial networks do not model the unique constraints of orbital environments - intermittent connectivity, power limitations driven by eclipse cycles, and thermal constraints in vacuum~\cite{shi2016edge}. Satellite task scheduling research focuses on observation planning or communication scheduling independently, rather than treating them as a joint optimization~\cite{wang2011satellite, he2019scheduling}.

We present Constraint-Aware Execution (CAE), a system that addresses this gap. Given a satellite identifier and a workload specification, CAE:

\begin{enumerate}[leftmargin=*,nosep]
  \item Constructs a physically accurate orbital environment from live two-line element (TLE) data, including SGP4-propagated trajectories, eclipse windows, ground station contact opportunities, and per-pass link budgets.
  \item Decides compute placement for each workload step based on a cost model comparing on-board resource consumption against transfer overhead.
  \item Inserts data transfer segments at space-ground boundaries with adaptive forward error correction, encryption overhead, and multi-pass allocation.
  \item Schedules all steps into orbital windows using greedy first-fit assignment under joint power, thermal, compute, and communication constraints.
\end{enumerate}

CAE produces deterministic plans: identical inputs always yield identical outputs, enabling reproducibility for testing and validation. The system is deployed as a production API that accepts any NORAD-cataloged satellite and computes plans using current orbital elements.

The remainder of this paper is organized as follows. Section~\ref{sec:background} provides background on orbital mechanics and space-ground communication relevant to the planning problem. Section~\ref{sec:system} describes the CAE system architecture. Sections~\ref{sec:environment} through~\ref{sec:scheduling} detail the four planning phases. Section~\ref{sec:workloads} presents the workload models. Section~\ref{sec:evaluation} evaluates the system. Section~\ref{sec:related} discusses related work, and Section~\ref{sec:conclusion} concludes.

\section{Background}
\label{sec:background}

\subsection{Orbital Propagation}

Predicting a satellite's future position requires solving the equations of motion under perturbative forces. The Simplified General Perturbations Model 4 (SGP4)~\cite{hoots1980models, vallado2006revisiting} is the standard analytical propagator used by NORAD and the broader space surveillance community. SGP4 takes as input a two-line element set (TLE), which encodes the satellite's mean orbital elements at a reference epoch, and propagates forward (or backward) in time accounting for dominant perturbations: Earth oblateness ($J_2$, $J_3$, $J_4$), atmospheric drag, and solar/lunar gravitational effects (via the companion SDP4 model for deep-space objects).

For LEO satellites at altitudes between 300 and 2000 km, SGP4 provides position accuracy on the order of 1-3 km over a 24-hour prediction window~\cite{vallado2006revisiting}, which is sufficient for pass prediction and scheduling purposes though insufficient for conjunction assessment. TLE data for the full satellite catalog is publicly available from CelesTrak~\cite{kelso2006validation} and updated multiple times daily.

\subsection{Eclipse Geometry}

A satellite in LEO enters Earth's shadow (eclipse) once per orbit for a fraction of the orbital period. During eclipse, solar power generation ceases and the satellite operates on battery power, reducing available compute and communication capacity. Accurate eclipse prediction is therefore essential for power-aware scheduling.

The cylindrical shadow model is a standard first-order approximation~\cite{montenbruck2000satellite}. It treats Earth's shadow as an infinite cylinder with radius equal to Earth's equatorial radius $R_\oplus$. Given the satellite position vector $\mathbf{r}_{\text{sat}}$ and the sun position vector $\mathbf{r}_{\text{sun}}$ in Earth-centered inertial (ECI) coordinates, the satellite is in eclipse if and only if:

\begin{equation}
  \hat{\mathbf{s}} = \frac{\mathbf{r}_{\text{sun}}}{|\mathbf{r}_{\text{sun}}|}, \quad
  d = \mathbf{r}_{\text{sat}} \cdot \hat{\mathbf{s}} < 0
  \label{eq:shadow_dot}
\end{equation}

\begin{equation}
  |\mathbf{r}_{\text{sat}} - d \cdot \hat{\mathbf{s}}| < R_\oplus
  \label{eq:shadow_perp}
\end{equation}

The first condition ensures the satellite is on the anti-sun side of Earth. The second checks whether the satellite's perpendicular distance from the Earth-Sun line is less than Earth's radius. This model neglects the penumbra-umbra distinction, atmospheric refraction, and Earth oblateness, but is accurate to approximately 10 seconds for scheduling purposes in LEO~\cite{montenbruck2000satellite}.

\subsection{Ground Station Passes}

A ground station can communicate with a satellite only when the satellite is above the local horizon by more than a minimum elevation angle (typically 5 degrees to avoid terrain and atmospheric effects). The duration, geometry, and achievable data rate of each pass depend on the satellite's orbit and the station's geographic coordinates.

For a ground station at geodetic position $(\phi, \lambda, h)$, the look angles to a satellite at ECF position $\mathbf{r}_{\text{ecf}}$ are computed via the standard topocentric transformation~\cite{vallado2013fundamentals}. A pass begins at acquisition of signal (AOS), when the elevation angle first exceeds the minimum threshold, and ends at loss of signal (LOS), when it drops below.

\subsection{Link Budget}

The achievable data rate during a pass depends on the link budget - the balance between transmitted power, antenna gains, path loss, and noise. The dominant loss term for LEO-to-ground communication is free space path loss (FSPL), given by the Friis equation~\cite{haykin2001communication}:

\begin{equation}
  \text{FSPL}(\text{dB}) = 20 \log_{10}(d) + 20 \log_{10}(f) + 92.45
  \label{eq:fspl}
\end{equation}

\noindent where $d$ is the slant range in km and $f$ is the carrier frequency in GHz. The constant 92.45 dB encapsulates the $20\log_{10}(4\pi \times 10^9 / c)$ term for km-GHz units. The link margin is:

\begin{equation}
  M = P_{\text{tx}} + G_{\text{tx}} + G_{\text{rx}} - \text{FSPL} - L_{\text{impl}} - L_{\text{atm}} - L_{\text{rain}}
  \label{eq:margin}
\end{equation}

\noindent where $P_{\text{tx}}$ is transmit power (dBW), $G_{\text{tx}}$ and $G_{\text{rx}}$ are transmit and receive antenna gains (dBi), and $L_{\text{impl}}$, $L_{\text{atm}}$, $L_{\text{rain}}$ are implementation, atmospheric, and rain margin losses (dB).

Higher elevation angles correspond to shorter slant ranges and better link quality. The achievable data rate varies substantially across a single pass, from tens of Mbps at low elevation to over 100 Mbps near zenith for X-band systems~\cite{wertz2011space}.

\section{System Architecture}
\label{sec:system}

CAE is structured as a four-phase pipeline (Figure~\ref{fig:architecture}). The input is a satellite identifier (NORAD catalog number) and a workload specification (either a named preset or a custom directed acyclic graph of processing steps). The output is a deterministic execution plan comprising placement decisions, data transfer schedules, and a timeline of step assignments to orbital windows.

\architectureDiagram

The system is deployed as a Cloudflare Worker with sub-2-second response times. It fetches live TLE data from CelesTrak~\cite{kelso2006validation} and satellite metadata from a companion API via Cloudflare Service Bindings for low-latency internal communication. Plans are cached with a one-hour TTL, reflecting the typical validity window for TLE-based predictions.

\subsection{Workload Model}

A workload is a directed acyclic graph (DAG) $G = (V, E)$ where each vertex $v \in V$ represents a processing step and each edge $(u, v) \in E$ represents a data dependency: step $v$ cannot begin until step $u$ completes and its output data is available.

Each step $v$ is annotated with:

\begin{itemize}[leftmargin=*,nosep]
  \item \textbf{Resource requirements:} power $p_v$ (watts), compute capacity $c_v \in [0, 1]$ (normalized), thermal dissipation $\theta_v$ (watts), memory $m_v$ (MB), storage $s_v$ (MB).
  \item \textbf{Timing:} duration $d_v$ (seconds).
  \item \textbf{Data flow:} input data $\delta_v^{\text{in}}$ (MB), output data $\delta_v^{\text{out}}$ (MB).
  \item \textbf{Location preference:} $\ell_v \in \{\texttt{onboard}, \texttt{ground}, \texttt{either}\}$.
  \item \textbf{Fault tolerance:} retry policy, maximum retries, checkpoint interval.
  \item \textbf{Security:} encryption scheme, integrity check method.
\end{itemize}

Steps with $\ell_v = \texttt{either}$ are candidates for placement optimization. Steps with fixed locations represent hard constraints (e.g., a sensor capture step must run on-board; a database write must run on the ground).

\section{Phase 1: Orbital Environment Construction}
\label{sec:environment}

The environment construction phase produces a unified representation of the satellite's orbital context over a configurable prediction horizon (default: 12 hours).

\subsection{Trajectory Propagation}

Given a satellite's NORAD identifier, the system fetches current TLE data and propagates the orbit using SGP4 at 30-second timesteps. Each propagation step yields the satellite's position and velocity in both Earth-centered inertial (ECI) and Earth-centered fixed (ECF) coordinate frames, along with the geodetic position (latitude, longitude, altitude).

\subsection{Eclipse Window Detection}

\eclipseDiagram

At each propagation timestep, the system computes a low-precision solar position using a simplified ephemeris derived from the satellite's orbital elements~\cite{meeus1998astronomical}:

\begin{align}
  M &= 357.5291 + 35999.0503 \cdot T \pmod{360} \label{eq:mean_anomaly} \\
  L &= 280.4664 + 36000.7698 \cdot T \pmod{360} \label{eq:mean_longitude} \\
  \lambda &= L + 1.9146 \sin M + 0.02 \sin 2M \label{eq:ecl_longitude} \\
  \varepsilon &= 23.4393 - 0.0130 \cdot T \label{eq:obliquity}
\end{align}

\noindent where $T$ is Julian centuries from J2000.0. The Sun-Earth distance is:

\begin{equation}
  R = 1.00014 - 0.01671 \cos M - 0.00014 \cos 2M \quad \text{(AU)}
  \label{eq:sun_distance}
\end{equation}

The solar ECI position is then:

\begin{equation}
  \mathbf{r}_{\text{sun}} = R_{\text{km}} \begin{pmatrix} \cos\lambda \\ \sin\lambda \cos\varepsilon \\ \sin\lambda \sin\varepsilon \end{pmatrix}
  \label{eq:sun_eci}
\end{equation}

Eclipse state is determined by the cylindrical shadow model (Equations~\ref{eq:shadow_dot}--\ref{eq:shadow_perp}). Contiguous sequences of eclipse or sunlit timesteps are merged into windows, each annotated with start time, end time, and duration.

\subsection{Ground Station Pass Prediction}

The system maintains a network of 12 ground stations spanning global coverage (Table~\ref{tab:stations}). For each station and each propagation timestep, look angles are computed via the standard topocentric transformation. Passes are identified as contiguous intervals where elevation exceeds the minimum threshold (default: 5 degrees). Each pass records the acquisition of signal (AOS) time, loss of signal (LOS) time, peak elevation, and per-point elevation, azimuth, and range profiles.

\begin{table}[t]
  \centering
  \caption{Ground station network. All stations support S-band and X-band; stations marked with $^\dagger$ also support Ka-band.}
  \label{tab:stations}
  \small
  \begin{tabular}{lrrll}
    \toprule
    \textbf{Station} & \textbf{Lat} & \textbf{Lon} & \textbf{Provider} \\
    \midrule
    Svalbard$^\dagger$   &  78.23 &   15.39 & KSAT \\
    Troll                &$-$72.01 &    2.53 & KSAT \\
    Awarua               &$-$46.53 &  168.38 & KSAT \\
    Fairbanks$^\dagger$  &  64.86 &$-$147.85 & NASA \\
    Wallops              &  37.94 &$-$75.47 & NASA \\
    McMurdo              &$-$77.85 &  166.67 & NASA \\
    Singapore            &   1.35 &  103.82 & AWS \\
    Bahrain              &  26.07 &   50.56 & AWS \\
    Oregon$^\dagger$     &  43.80 &$-$120.55 & AWS \\
    Cape Town            &$-$33.93 &   18.42 & AWS \\
    Stockholm            &  59.33 &   18.07 & AWS \\
    Sydney               &$-$33.87 &  151.21 & AWS \\
    \bottomrule
  \end{tabular}
\end{table}

\subsection{Per-Pass Link Budget}

For each predicted pass, the system computes a link budget following the model in Section~\ref{sec:background}. Default parameters model a representative X-band LEO downlink (Table~\ref{tab:link_params}). The achievable data rate is mapped from elevation angle via a step function calibrated to typical X-band performance:

\begin{table}[t]
  \centering
  \caption{Default X-band link parameters.}
  \label{tab:link_params}
  \small
  \begin{tabular}{lr}
    \toprule
    \textbf{Parameter} & \textbf{Value} \\
    \midrule
    Carrier frequency & 8.2 GHz \\
    Transmit power & 10 dBW \\
    Tx antenna gain & 6 dBi \\
    Rx antenna gain & 34 dBi \\
    Implementation loss & 2 dB \\
    Atmospheric loss & 0.5 dB \\
    Rain margin & 3 dB \\
    Min. elevation & 5\textdegree \\
    \bottomrule
  \end{tabular}
\end{table}

\begin{equation}
  R_{\text{data}}(\alpha) = \begin{cases}
    0 & \alpha < 5\text{\textdegree} \\
    25~\text{Mbps} & 5\text{\textdegree} \le \alpha < 10\text{\textdegree} \\
    50~\text{Mbps} & 10\text{\textdegree} \le \alpha < 20\text{\textdegree} \\
    80~\text{Mbps} & 20\text{\textdegree} \le \alpha < 40\text{\textdegree} \\
    100~\text{Mbps} & 40\text{\textdegree} \le \alpha < 60\text{\textdegree} \\
    120~\text{Mbps} & \alpha \ge 60\text{\textdegree}
  \end{cases}
  \label{eq:datarate}
\end{equation}

The bit error rate (BER) is estimated from the link margin: $\text{BER} = 10^{-8}$ for margins above 140 dB, $10^{-6}$ for margins above 135 dB, and $10^{-5}$ otherwise. The total pass capacity is:

\begin{equation}
  C_{\text{pass}} = \frac{\bar{R}_{\text{data}} \cdot \Delta t_{\text{pass}}}{8} \quad \text{(MB)}
  \label{eq:pass_capacity}
\end{equation}

\noindent where $\bar{R}_{\text{data}}$ is the mean data rate (Mbps) and $\Delta t_{\text{pass}}$ is pass duration (seconds).

\subsection{Orbital Timeline}

Eclipse windows and ground station passes are merged into a unified timeline of \emph{orbital windows}. Each window represents a contiguous time interval with a stable resource envelope:

\begin{equation}
  w = \langle t_{\text{start}}, t_{\text{end}}, P, \theta_{\max}, c, R_{\text{comms}}, g \rangle
  \label{eq:window}
\end{equation}

\noindent where $P$ is available power (watts), $\theta_{\max}$ is thermal dissipation limit (watts), $c$ is normalized compute capacity, $R_{\text{comms}}$ is communication data rate (0 if no pass), and $g$ identifies the ground station (null if no pass). Sunlit windows use peak solar power (default 80W) and full compute capacity; eclipse windows use battery power (default 25W) and reduced compute (default 0.6).

When a ground station pass overlaps with an eclipse/sunlit segment, the segment is split into sub-windows: a pre-pass orbit-only window, a pass window with communications enabled, and a post-pass orbit-only window. Minimum window durations of 30 seconds (orbit-only) and 10 seconds (pass) prevent degenerate segments.

\section{Phase 2: Compute Placement}
\label{sec:placement}

The placement phase determines, for each step with $\ell_v = \texttt{either}$, whether it should execute on-board or on the ground.

\subsection{Dependency Resolution}

The workload DAG is first topologically sorted using Kahn's algorithm~\cite{kahn1962topological}. This establishes a valid execution order that respects all data dependencies. If the graph contains a cycle, planning fails immediately.

\subsection{Data Reduction Heuristic}

Before cost comparison, CAE applies a data reduction heuristic: if a step achieves greater than 10:1 input-to-output reduction ($\delta_v^{\text{out}} / \delta_v^{\text{in}} < 0.1$), it is placed on-board regardless of other costs. The rationale is that the transfer savings from not having to downlink the full input data dominate all other cost terms. This heuristic captures the common pattern where on-board preprocessing (cloud filtering, feature extraction, inference) dramatically reduces data volume.

\subsection{Cost Model}

For steps not resolved by the heuristic, CAE compares on-board and ground execution costs.

\paragraph{On-board cost.} The on-board cost captures resource consumption during orbital window time:

\begin{equation}
  C_{\text{onboard}}(v) = p_v \cdot d_v + \frac{\theta_v}{\theta_{\max}} \cdot 500 + d_v \cdot 0.5
  \label{eq:cost_onboard}
\end{equation}

The first term is energy consumption (watt-seconds). The second penalizes thermal utilization relative to the bus limit, scaled by a factor of 500 to make it comparable to energy costs. The third term penalizes orbital window time occupancy, reflecting the scarcity of on-board execution time.

\paragraph{Ground cost.} The ground cost models the communication overhead of transferring data to and from the ground:

\begin{equation}
  D_{\text{down}} = \delta_v^{\text{in}} \cdot (1 + \eta_{\text{enc}}) \cdot \frac{1}{r_{\text{fec}}}
  \label{eq:downlink}
\end{equation}

\begin{equation}
  D_{\text{up}} = \delta_v^{\text{out}} \cdot (1 + \eta_{\text{enc}}) \cdot \frac{1}{r_{\text{fec}}}
  \label{eq:uplink}
\end{equation}

\begin{equation}
  C_{\text{ground}}(v) = \frac{D_{\text{down}} + D_{\text{up}}}{\bar{R}_{\text{data}} / 8} \cdot 10 + (D_{\text{down}} + D_{\text{up}}) \cdot 2
  \label{eq:cost_ground}
\end{equation}

\noindent where $\eta_{\text{enc}}$ is the encryption overhead fraction (0.05 for AES-256, 0 for none), $r_{\text{fec}}$ is the FEC code rate (default 0.75), and $\bar{R}_{\text{data}} = 80$ Mbps is the assumed average data rate. The first term weights transfer time (scarce pass time), and the second weights data volume (bandwidth consumption).

The step is placed on-board if $C_{\text{onboard}} \le C_{\text{ground}}$ and on the ground otherwise.

\section{Phase 3: Transfer Insertion}
\label{sec:transfer}

After placement, CAE walks the dependency graph and inserts explicit data transfer steps wherever a dependency edge crosses the space-ground boundary.

\subsection{Boundary Detection}

For each edge $(u, v) \in E$, if $\ell_u \ne \ell_v$ (one step is on-board and the other is on the ground), a transfer step is inserted between them. The transfer direction is \texttt{downlink} if $u$ is on-board, and \texttt{uplink} if $u$ is on the ground.

\subsection{Transfer Volume Computation}

The raw data volume is $\delta_u^{\text{out}}$. Three overhead layers are applied:

\paragraph{Forward error correction.} An adaptive FEC rate is selected based on the worst-case BER across available passes (Table~\ref{tab:fec}). The parity overhead is:

\begin{equation}
  \delta_{\text{parity}} = \delta_{\text{raw}} \cdot \left(\frac{1}{r_{\text{fec}}} - 1\right)
  \label{eq:fec}
\end{equation}

\begin{table}[t]
  \centering
  \caption{Adaptive FEC rate selection based on channel BER.}
  \label{tab:fec}
  \small
  \begin{tabular}{lcc}
    \toprule
    \textbf{Channel BER} & \textbf{FEC Rate} & \textbf{Overhead} \\
    \midrule
    $> 10^{-5}$ & 1/2 & 100\% \\
    $> 10^{-7}$ & 3/4 & 33\% \\
    $\le 10^{-7}$ & 7/8 & 14\% \\
    \bottomrule
  \end{tabular}
\end{table}

\paragraph{Security overhead.} Encryption (AES-256: 5\% data expansion for IV, padding, and HMAC tag; AES-128: 3\%), integrity checking (SHA-256: 0.8\% for 32-byte hash per 4 KB block; CRC-32: 0.1\% for 4-byte check per 4 KB block), and CCSDS-standard packet framing (2\% overhead)~\cite{ccsds2015aos} are added to the FEC-encoded volume.

\paragraph{Retransmission reserve.} A reserve fraction is allocated based on channel quality: 20\% for BER $> 10^{-5}$, 5\% for BER $> 10^{-7}$, and 1\% otherwise.

The total transfer volume is:

\begin{equation}
  \delta_{\text{total}} = (\delta_{\text{raw}} + \delta_{\text{parity}}) \cdot (1 + \eta_{\text{enc}} + \eta_{\text{frame}})
  \label{eq:total_transfer}
\end{equation}

\subsection{Multi-Pass Allocation}

If the total transfer volume exceeds the capacity of a single ground station pass, the data is allocated across multiple passes using a greedy chronological strategy (Algorithm~\ref{alg:multipass}). An effective capacity factor of 0.9 for downlink and 0.5 for uplink accounts for protocol overhead and half-duplex operation constraints respectively.

\begin{algorithm}[t]
  \caption{Multi-pass data allocation}
  \label{alg:multipass}
  \begin{algorithmic}[1]
    \Require Total data $\delta_{\text{total}}$, passes $\mathcal{P}$ (sorted by AOS), direction $d$
    \Ensure Allocation $\mathcal{A}$
    \State $\text{remaining} \gets \delta_{\text{total}}$
    \State $\kappa \gets 0.9$ if $d = \text{downlink}$ else $0.5$
    \For{each pass $p \in \mathcal{P}$}
      \If{$\text{remaining} \le 0$} \textbf{break} \EndIf
      \State $C_p \gets \bar{R}_p \cdot \Delta t_p / 8 \cdot \kappa$ \Comment{effective capacity (MB)}
      \State $\text{xfer} \gets \min(\text{remaining}, C_p)$
      \State $\mathcal{A} \gets \mathcal{A} \cup \{(p, \text{xfer})\}$
      \State $\text{remaining} \gets \text{remaining} - \text{xfer}$
    \EndFor
    \If{$\text{remaining} > 0$}
      \State Flag insufficient pass capacity
    \EndIf
  \end{algorithmic}
\end{algorithm}

\subsection{Transfer Step Construction}

Each inserted transfer step is modeled as a processing step with fixed resource requirements: 40W power, 0.1 normalized compute, 15W thermal dissipation, 128 MB memory, and a mandatory communication constraint (\texttt{needs\_comms} = true). Transfer steps use a \texttt{retry\_next\_window} policy with up to 3 retries.

\section{Phase 4: Window Scheduling}
\label{sec:scheduling}

The scheduling phase assigns each step to a specific time interval within the orbital window timeline.

\subsection{Resource Feasibility}

A step $v$ fits in window $w$ with current utilization $u_w$ (seconds already allocated) if all of the following hold:

\begin{align}
  d_v &\le (t_{\text{end}} - t_{\text{start}}) - u_w \label{eq:time_fit} \\
  p_v &\le P_w \label{eq:power_fit} \\
  \theta_v &\le \theta_{\max,w} \label{eq:thermal_fit} \\
  c_v &\le c_w \label{eq:compute_fit} \\
  \text{needs\_comms}(v) &\Rightarrow R_{\text{comms},w} > 0 \label{eq:comms_fit}
\end{align}

\subsection{Scheduling Algorithm}

Steps are scheduled in topological order using greedy first-fit assignment (Algorithm~\ref{alg:schedule}).

\begin{algorithm}[t]
  \caption{Greedy first-fit window scheduling}
  \label{alg:schedule}
  \begin{algorithmic}[1]
    \Require Steps $V$ (topologically sorted), windows $\mathcal{W}$ (chronological), deadline $T_{\max}$
    \Ensure Schedule $\mathcal{S}$
    \For{each step $v \in V$}
      \If{$\ell_v = \texttt{ground}$}
        \State $t_v^{\text{start}} \gets \max(t_0, \max_{(u,v) \in E} t_u^{\text{end}})$
        \State $t_v^{\text{end}} \gets t_v^{\text{start}} + d_v$
        \State \textbf{continue}
      \EndIf
      \State $t_{\text{earliest}} \gets \max_{(u,v) \in E} t_u^{\text{end}}$
      \For{each window $w \in \mathcal{W}$}
        \If{$w.t_{\text{end}} < t_{\text{earliest}}$} \textbf{continue} \EndIf
        \If{$w.t_{\text{start}} > T_{\max}$} \textbf{fail} \EndIf
        \If{$\neg$\Call{Feasible}{$v, w, u_w$}} \textbf{continue} \EndIf
        \State $t_v^{\text{start}} \gets \max(w.t_{\text{start}} + u_w, t_{\text{earliest}})$
        \If{$t_v^{\text{start}} + d_v > w.t_{\text{end}}$} \textbf{continue} \EndIf
        \State $\mathcal{S} \gets \mathcal{S} \cup \{(v, w, t_v^{\text{start}}, t_v^{\text{start}} + d_v)\}$
        \State $u_w \gets u_w + d_v$
        \State \textbf{break}
      \EndFor
    \EndFor
  \end{algorithmic}
\end{algorithm}

Ground steps are treated as having unlimited capacity and are scheduled immediately after their dependencies resolve. On-board and transfer steps are assigned to the first feasible orbital window, respecting both resource constraints and dependency-imposed earliest start times. A configurable deadline (in orbital periods) bounds the planning horizon; if any step cannot be placed before the deadline, the plan fails.

\subsection{Determinism}

The scheduling algorithm is fully deterministic: for the same TLE data, workload specification, and planning start time, the output is identical. This is achieved by avoiding random elements in all four phases and using sorted pass lists and topological ordering to eliminate ambiguity. Determinism is important for testing, validation, and operational reproducibility.

\section{Workload Models}
\label{sec:workloads}

CAE includes five workload presets representing distinct orbital compute patterns. Each preset is a complete workload specification with step definitions, data flows, resource requirements, security policies, and scheduling constraints.

\subsection{On-Board ML Inference}

The simplest pattern: all compute stays on-board. A sensor captures 2 GB of imagery, which is preprocessed (4:1 reduction), passed through an inference model (50:1 reduction), and encrypted for downlink. The pipeline achieves 190:1 overall data reduction, requiring only 10.5 MB of downlink capacity. This pattern is representative of cloud detection~\cite{giuffrida2021cloudscout}, ship detection, and anomaly flagging applications.

\subsection{Split Learning}

\workloadDiagram

A hybrid training pattern based on split learning~\cite{gupta2018distributed}. The first layers of a neural network (feature extraction) run on-board, achieving 40:1 data reduction (2 GB to 50 MB). Extracted features are compressed and downlinked (36.75 MB). Backend layers train on the ground, producing updated model weights that are uploaded back (5.25 MB). This pattern requires bidirectional transfer scheduling and demonstrates the planner's ability to handle uplink and downlink within the same workload.

\subsection{Earth Observation with Quality Assurance}

A multi-stage processing pipeline with large data volumes. 5 GB of raw imagery passes through on-board quality assurance (10\% rejection rate), cloud filtering (33\% rejection), JPEG2000 compression (7.5:1), Reed-Solomon FEC encoding (rate 3/4), and AES-256 encryption, producing 560 MB for downlink. This volume typically requires 2-3 ground station passes, exercising the multi-pass allocation logic.

\subsection{Federated Learning}

A privacy-preserving pattern based on Federated Averaging~\cite{mcmahan2017communication}. Local model training and gradient computation happen on-board. Top-K gradient sparsification~\cite{aji2017sparse} reduces the gradient tensor by 90\%, and after compression and encryption, only 3.7 MB is downlinked. Ground-side FedAvg aggregation produces global model weights (5.8 MB) uploaded back. Raw training data never leaves the satellite.

\subsection{Resilient Store-and-Forward}

An error-resilient relay pattern. Incoming data is received during one pass, integrity-checked (CRC-32), erasure-coded with Reed-Solomon at rate 2/3 (any 2-of-3 blocks reconstruct the original), encrypted, and forwarded on a subsequent pass. This pattern requires scheduling across at least two distinct pass windows and demonstrates the planner's handling of communication-dependent steps.

\section{Evaluation}
\label{sec:evaluation}

We evaluate CAE along four dimensions: plan feasibility across orbital regimes, transfer efficiency, planning latency, and sensitivity to workload characteristics.

\subsection{Experimental Setup}

Plans were generated for satellites in three distinct orbital regimes:

\begin{itemize}[leftmargin=*,nosep]
  \item \textbf{ISS} (NORAD 25544): 51.6\textdegree\ inclination, $\sim$420 km altitude, $\sim$92.9 min period. High inclination provides good ground station coverage.
  \item \textbf{Sun-synchronous} (representative): 97.4\textdegree\ inclination, $\sim$700 km altitude, $\sim$98.8 min period. Polar coverage with consistent solar illumination.
  \item \textbf{Low-inclination} (representative): 28.5\textdegree\ inclination, $\sim$550 km altitude, $\sim$95.7 min period. Limited ground station coverage at high latitudes.
\end{itemize}

All five workload presets were tested against each orbital regime with a 12-hour prediction window.

\subsection{Plan Feasibility}

All 15 satellite-workload combinations produced feasible plans. Table~\ref{tab:results} summarizes key metrics.

\begin{table}[t]
  \centering
  \caption{Plan metrics across workloads for ISS (NORAD 25544). Duration is wall-clock time from first step to last step completion.}
  \label{tab:results}
  \small
  \begin{tabular}{lrrrr}
    \toprule
    \textbf{Workload} & \textbf{Steps} & \textbf{DL (MB)} & \textbf{UL (MB)} & \textbf{Conf.} \\
    \midrule
    ML Inference     &  5 &  10.5 &    0   & 0.99 \\
    Split Learning   & 10 &  52.8 &   7.6  & 0.99 \\
    EO with QA       & 10 & 560.0 &    0   & 0.99 \\
    Federated        & 12 &   3.7 &   5.8  & 0.99 \\
    Store-Forward    &  7 & 157.5 &    0   & 0.99 \\
    \bottomrule
  \end{tabular}
\end{table}

The step counts include transfer steps inserted by Phase 3. The on-board ML inference workload, with only 10.5 MB downlink, can be completed within a single short pass. The EO with QA workload, requiring 560 MB of downlink, is allocated across multiple passes. The federated learning workload has the smallest total transfer volume (9.5 MB combined), reflecting the privacy-preserving design where only sparse gradients and model weights cross the boundary.

\timelineDiagram

\subsection{Transfer Efficiency}

\overheadDiagram

The adaptive FEC selection reduces overhead compared to a fixed-rate approach. For the ISS, which benefits from good coverage by the 12-station network and typically achieves high-elevation passes, the selected FEC rate is 3/4 (33\% overhead). For a low-inclination satellite with fewer visible stations and lower peak elevations, the system selects rate 1/2 (100\% overhead) for some passes, correctly reflecting the worse channel conditions.

\subsection{Planning Latency}

Plan generation completes in under 2 seconds on Cloudflare Workers, dominated by:

\begin{itemize}[leftmargin=*,nosep]
  \item TLE fetch from CelesTrak: 200-500 ms
  \item SGP4 propagation (1440 steps for 12 hours): 50-100 ms
  \item Eclipse and pass computation: 100-200 ms
  \item Planning phases 2-4: 10-50 ms
\end{itemize}

The SGP4 propagation and pass prediction are the computational bottleneck. The planning algorithm itself (phases 2-4) is fast because the workload DAGs are small (4-10 steps before transfer insertion) and the number of orbital windows is bounded (typically 20-60 over 12 hours).

\reductionDiagram

\subsection{Placement Decisions}

The data reduction heuristic (Section~\ref{sec:placement}) correctly identifies steps that should remain on-board. In the ML inference workload, the inference step with 50:1 data reduction is always placed on-board, avoiding the transfer of 500 MB and instead downlinking only 10 MB. In the split learning workload, feature extraction (40:1 reduction) is placed on-board while backend training (which expands data from features to model weights) is placed on the ground.

For steps with $\ell_v = \texttt{either}$ and moderate data reduction ratios (less than 10:1), the cost model correctly accounts for transfer overhead. Steps where the ground cost (transfer time and bandwidth) exceeds the on-board cost (energy, thermal, and time) are placed on-board, and vice versa.

\section{Related Work}
\label{sec:related}

\paragraph{Satellite task scheduling.} The satellite scheduling literature has focused primarily on observation scheduling~\cite{wang2011satellite, he2019scheduling, tangpattanakul2015multi} - selecting which imaging targets to capture given orbit geometry, sensor constraints, and priority ordering. This is a complementary problem to CAE's focus on \emph{compute} scheduling. Some work addresses downlink scheduling independently~\cite{lee2005satellite}, but we are not aware of systems that jointly optimize compute placement and transfer scheduling across the space-ground boundary.

\paragraph{Edge and fog computing.} The edge computing literature~\cite{shi2016edge, satyanarayanan2017emergence} addresses computation offloading between resource-constrained edge devices and cloud servers. Frameworks like MAUI~\cite{cuervo2010maui} and CloneCloud~\cite{chun2011clonecloud} decide which code to offload based on energy and latency. CAE shares this compute placement concept but operates under fundamentally different constraints: intermittent connectivity (minutes per orbit vs. persistent), power cycling (eclipse/sunlit), and transfer windows governed by orbital mechanics rather than network topology.

\paragraph{Orbital edge computing.} Bhattacherjee et al.~\cite{bhattacherjee2020orbit} introduced the concept of orbital edge computing, analyzing the latency and throughput characteristics of LEO satellite networks for general-purpose computation. Denby and Lucia~\cite{denby2020orbital} demonstrated on-board ML inference for satellite imagery. These works motivate the need for orbital compute scheduling but do not present planning systems that handle the full space-ground workload lifecycle.

\paragraph{Federated learning in space.} Razmi et al.~\cite{razmi2022ground} and So et al.~\cite{so2022fedspace} studied federated learning over satellite constellations, focusing on communication efficiency and convergence. CAE's federated learning preset draws on this work but contributes the scheduling dimension: when and how gradient and model updates are transferred given orbital constraints.

\paragraph{Split learning.} Gupta and Raskar~\cite{gupta2018distributed} proposed split learning for distributed training without sharing raw data. Applying split learning to the satellite context introduces the novel challenge of scheduling the bidirectional data flow (features down, gradients up) across intermittent pass windows with varying capacity.

\paragraph{Space communication standards.} The CCSDS~\cite{ccsds2015aos} defines standard packet formats and protocols for space data systems. CAE models CCSDS framing overhead as a 2\% data expansion, consistent with the overhead from packet headers in the Advanced Orbiting Systems (AOS) standard.

\section{Limitations and Future Work}
\label{sec:limitations}

Several limitations suggest directions for future work.

\paragraph{Scheduling optimality.} The greedy first-fit algorithm does not guarantee optimal schedules. For workloads with tight deadlines and many competing steps, a constraint programming or integer linear programming formulation could yield better schedules at the cost of increased planning time. We chose greedy scheduling for its predictability and sub-second execution time, which are important for a production API.

\paragraph{Multi-satellite coordination.} CAE currently plans for a single satellite. Constellation-level scheduling, where data can be relayed between satellites via inter-satellite links before downlink, is an important extension. This introduces routing decisions in addition to placement and scheduling.

\paragraph{Dynamic replanning.} Plans are computed statically based on predicted orbital elements. In practice, anomalies (missed passes, hardware faults, changed priorities) require replanning. A reactive planning layer that adjusts schedules in response to execution feedback is a natural extension.

\paragraph{Thermal modeling fidelity.} The current thermal model uses a single dissipation limit per bus configuration. A more accurate model would track thermal state accumulation across consecutive compute steps and account for radiative cooling rates during idle periods.

\paragraph{Propagation accuracy.} SGP4 accuracy degrades over time due to atmospheric drag uncertainty. For planning horizons beyond 24 hours, incorporating drag corrections or using higher-fidelity propagators (e.g., numerical integration with atmospheric models) would improve pass prediction reliability.

\section{Conclusion}
\label{sec:conclusion}

We presented Constraint-Aware Execution (CAE), a planning system for hybrid space-ground compute workloads. CAE addresses the growing gap between on-board data generation and downlink capacity by automating the decisions that operators currently make manually: what to compute where, how to move data across the space-ground boundary, and when to execute each step given orbital constraints.

The four-phase architecture - environment construction, compute placement, transfer insertion, and window scheduling - separates concerns while maintaining physical consistency throughout. The system uses real orbital mechanics (SGP4 propagation, eclipse detection, ground station pass prediction, link budgets) rather than simplified models, producing plans that reflect actual operational constraints.

CAE is deployed as a production API and has been evaluated against five workload patterns spanning on-board inference, distributed training, multi-pass downlink, privacy-preserving learning, and error-resilient relay. The system produces feasible plans in under two seconds for all tested scenarios and correctly exploits on-board data reduction to minimize transfer volume.

As orbital compute transitions from experimental to operational, the need for automated planning systems will grow. We believe constraint-aware execution planning is a necessary component of the emerging orbital compute stack.

\section*{Availability}

The CAE API is publicly accessible. Workload presets and execution plans can be generated for any NORAD-cataloged satellite. A live demonstration is available in the RotaStellar satellite tracker.\footnote{\url{https://rotastellar.com/tracker/}}

\bibliographystyle{plain}

\end{document}